\documentclass[lettersize,journal]{IEEEtran}
\usepackage{amsmath,amsfonts}
\usepackage{algorithmic}
\usepackage{algorithm}
\usepackage{array}
\usepackage{subfig}
\usepackage{textcomp}
\usepackage{stfloats}
\usepackage{url}
\usepackage{verbatim}
\usepackage{graphicx}
\usepackage{cite}
\usepackage{mathtools}
\usepackage{booktabs}
\usepackage{amssymb}
\usepackage{lettrine}
\usepackage{color}
\usepackage{multirow}
\usepackage{hyperref}
\graphicspath{{./figure/}}
\begin{document}

\title{Revisiting General Map Search via Generative \\ Point-of-Interest Retrieval}

\author{Dong~Chen,
        Shuai~Zheng,
        Haoyang~Shao,
        Hongsheng~Wu,
        Muhao~Xu,
        Yeyu~Yan,
        Ruifang~Li,
        and~Zhenfeng~Zhu$^{*}$
\thanks{$^{*}$Corresponding author: Zhenfeng Zhu.}
\thanks{Dong Chen, Shuai Zheng, Muhao Xu, Yeyu Yan, and Zhenfeng Zhu are with the Institute of Information Science, Beijing Jiaotong University, Beijing 100044, China (e-mail: dchen2001@bjtu.edu.cn; zs1997@bjtu.edu.cn; mhxu1998@bjtu.edu.cn; 23111086@bjtu.edu.cn; zhfzhu@bjtu.edu.cn).}
\thanks{Haoyang Shao, Hongsheng Wu, and Ruifang Li are with the Map Platform Department, Tencent Inc., China (e-mail: howieshao@tencent.com; hongshengwu@tencent.com; finali@tencent.com).}
}

\markboth{Journal of \LaTeX\ Class Files,~Vol.~14, No.~8, August~2021}%
{Chen \MakeLowercase{\textit{et al.}}: Revisiting General Map Search via Generative POI Retrieval}

\maketitle

\begin{abstract}
Point-of-Interest (POI) retrieval aims to identify relevant candidates from massive-scale POI databases, serving as a cornerstone for diverse location-based services. However, in general map search scenarios, conventional POI retrieval methods are increasingly challenged by underspecified user queries due to their excessive reliance on surface-level semantic matching. Meanwhile, such queries are often highly context-dependent and personalized, yet existing retrieval paradigms struggle to effectively synergize heterogeneous contexts for complex search intent inference. To address these limitations, we revisit general map search from a generative perspective and propose \textbf{GenPOI}, an innovative \underline{\textbf{Gen}}erative \underline{\textbf{POI}} retrieval framework tailored for general search on maps. It seamlessly unifies heterogeneous search contexts and POIs into structured sequences, leveraging the powerful contextual modeling of Large Language Models (LLMs) for spatial-aware candidate generation. Consequently, this generative paradigm effectively solves more challenging queries through profound context dependency modeling and search intent reasoning. Specifically, accounting for the unique geospatial nature of map scenarios, GenPOI introduces a novel \textit{Geo-Semantic POI Tokenization} to represent each POI as a compact token sequence encoding both semantic and geographic context, thus grounding the LLM's spatial understanding. Additionally, a \textit{proximity-aware constrained generation} strategy is employed to restrict the decoding space of the LLM, ensuring the validity and geospatial relevance of the generated results. Extensive experiments on large-scale industrial datasets from Tencent Map, comprising POIs at the scale of \textbf{over 10 million}, demonstrate the superior performance of GenPOI.
\end{abstract}


\section{Introduction}
\IEEEPARstart{W}{ith} the pervasiveness of location-based services (LBS)~\cite{LBSbook}, map search has emerged as a critical gateway for people to access spatial information and explore points-of-interest (POIs).  As a cornerstone of map search scenarios, POI retrieval aims to identify a set of relevant candidates from large-scale POI databases based on users' queries, locations, and other search context. Therefore, the quality of retrieval results, which encompasses both semantic and spatial relevance, directly influences users' decision-making efficiency and satisfaction. Early POI retrieval methods were grounded in basic lexical or prefix matching~\cite{bm25, keywordRetrieval, wordembeddingRetrieval} techniques built on textual POI metadata (e.g., POI names, addresses, and category labels). Driven by the demand for more precise semantic-level retrieval, recent advances have steered the paradigm toward neural embedding models~\cite{DSSM, neural_embedding, neural_embedding_2}, which learn dense representations for both queries and POIs to quantify their relevance in a shared semantic space. To better adapt to map search scenarios, these models~\cite{PALM, DPSM, PR-geo2} typically incorporate geographical considerations through textual location descriptions, dedicated geo-encoders, or direct multi-stage geographic constraints, thereby integrating spatial signals into the POI retrieval pipeline.

\begin{figure}[ht]
  \centering
  \includegraphics[width=\columnwidth]{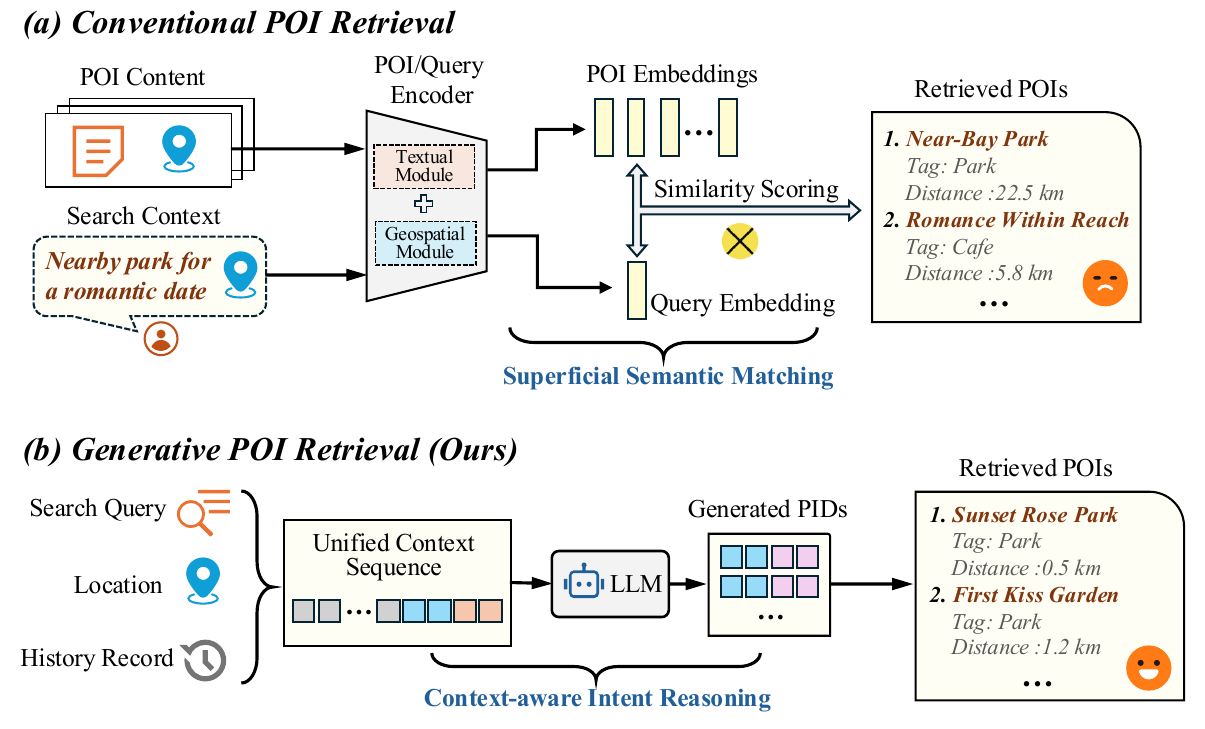}
  \caption{Comparison of different POI retrieval paradigms. Unlike superficial semantic matching, our generative paradigm unifies diverse search contexts and leverages LLMs for sophisticated intent reasoning, thereby effectively handling general search queries. }
  \label{intro}
\end{figure}

In general map search scenarios, users are increasingly issuing more underspecified and complex queries (e.g., ``nearby park for a romantic date'') that are often only confined to category-level semantic constraints. These queries are highly exploratory: the user's intent is not to locate a specific POI but rather to discover suitable, personalized options that align with their complex search context. However, as shown in Fig.~\ref{intro}(a), current retrieval frameworks remain anchored to the paradigm of surface-level lexical or semantic matching, making them ill-equipped to handle such challenging general search queries. Furthermore, existing methods struggle to seamlessly integrate heterogeneous contextual signals, such as geographical coordinates, textual queries, and user profiles, resulting in fragmented feature fusion. The absence of a unified representation prevents the model from capturing cross-modal dependencies among these distinct inputs, thereby limiting their ability to perform sophisticated intent reasoning.

Driven by the rapid advancements in Large Language Models (LLMs)~\cite{LLMSurvey1,LLMSurvey2}, generative retrieval (GR)~\cite{GRsurvey} has emerged as a promising paradigm, demonstrating superior performance in tasks demanding high contextual awareness. Unlike traditional discriminative paradigms, these methods encode entities into discrete semantic IDs (SIDs)~\cite{tiger}, reformulating retrieval as an SID token generation task. These SIDs are compact and structured relative to the original complex attributes, allowing LLMs to more readily learn the relational correlations between entities, queries, and contextual information. Building on this, GR leverages the inherent reasoning capabilities of LLMs to decode complex sequential contexts, thus enabling the retrieval of candidates that align precisely with implicit user intent. Nevertheless, most existing GR methods are primarily designed for recommender systems~\cite{OneRec, poirecsuervey} or document retrieval~\cite{GR4Doc,GR4Doc-2}, where the objective is to generate target entities conditioned solely on historical behaviors or purely textual inputs. Accordingly, their lack of geospatial awareness limits their adaptability to map search scenarios, where spatial and semantic aspects must be jointly considered.

In practice, designing an elegant generative retrieval framework for POIs is non-trivial, as it faces two primary limitations. \textbf{(1) Geospatial Perception Gap}: Unlike other types of entities, POIs are anchored to extrinsic geographic locations. Bridging this gap necessitates a specialized approach that empowers LLMs to effectively perceive spatial relationships between POIs and users within a discretized token space.
\textbf{(2) Hallucination in Vast Generation Spaces}: Due to the expansive vocabulary of LLMs and the massive scale of POI data, the LLM is prone to generating hallucinated results. This necessitates explicit generation constraints to ensure both the validity and geospatial reasonableness of the retrieved POIs.

Given these limitations, we propose \textbf{GenPOI}, a POI-oriented generative retrieval framework that explicitly incorporates map-specific geospatial characteristics. As illustrated in Fig.~\ref{fw}(b), by linearizing the textual queries, geographical locations, POIs, and historical user behaviors into a unified context sequence, GenPOI harnesses the powerful contextual modeling capabilities of LLMs to facilitate geographically-aware target POI generation. In detail, GenPOI introduces a novel \textbf{Geo-Semantic POI Tokenization} to harmonize POI entities with the LLM's discrete linguistic space. It encodes each POI into a compact POI ID (PID), comprising structured discrete tokens that integrate both semantic attributes and geographic locations. By replacing numerical atomic IDs with these interpretable PIDs, the LLM is enabled to readily comprehend a POI by directly perceiving ``\textit{what it is}'' and ``\textit{where it is}'' within a unified vocabulary. Considering the expansive generation space of LLMs, \textbf{proximity-aware constrained generation} is introduced to impose spatial and vocabulary constraints during POI generation. This mechanism ensures both the validity and efficiency of PID generation, effectively preventing the hallucination of non-existent and geospatially unreasonable POIs. To evaluate the effectiveness of the proposed model, we conduct extensive experiments on large-scale industrial map search datasets, with the POI scale exceeding \textbf{10 million}. Our main contributions can be summarized as follows:
\begin{itemize}
    \item We propose \textbf{GenPOI}, a novel generative POI retrieval framework that unifies heterogeneous search contexts and leverages the powerful contextual modeling capabilities of LLMs to significantly enhance retrieval performance in general search scenarios.
    \item To bridge discrete POI entities and the LLM's linguistic space, a novel Geo-Semantic POI Tokenization module is introduced to enhance the LLM's semantic and spatial understanding for POIs.
    \item Considering the vast generation space of LLMs, a proximity-aware constrained generation strategy is designed to ensure the validity and spatial plausibility of the generated results.
    \item Extensive experiments on real-world industrial datasets demonstrate that GenPOI significantly outperforms state-of-the-art baselines in general search scenarios.
\end{itemize}

\section{Related Work}
\subsection{POI Retrieval}
POI retrieval aims to identify a set of relevant candidates from a large database that match a user's query. Early methods \cite{keywordRetrieval,wordembeddingRetrieval} relied on exact term matching of textual information, and thus often suffered from the vocabulary mismatch problem, in which semantically similar words are treated as unrelated. With the development of deep learning, mainstream POI retrieval methods have shifted to neural embedding-based semantic matching~\cite{DSSM,DSSM-LSTM}. These approaches leverage Deep Neural Networks (DNNs), particularly Pre-trained Language Models (PLMs)~\cite{pretrain-1,pretrain-2,bge-m3}, to map queries and POIs into a shared latent space, quantifying relevance through vector similarity~\cite{neural_embedding,neural_embedding_2}.  To account for the spatial nature of POIs, recent frameworks often integrate spatial context~\cite{PALM, PR-geo2}, ranging from textualized address descriptions and dedicated geographic encoders to explicit spatial constraints. Furthermore, state-of-the-art methods~\cite{hgamn,PR-geo3,PR-userprofile} incorporate more complex heterogeneous contexts including temporal signals, historical behaviors, and user profiles to enhance retrieval performance.

Despite these advancements, conventional approaches often struggle to resolve the inherent ambiguity of queries in general map search scenarios. Furthermore, they face a critical bottleneck in seamlessly unifying heterogeneous search contexts, which hinders their ability to perform sophisticated search intent reasoning.

\begin{figure*}[ht]
  \centering
  \includegraphics[width=0.8\textwidth]{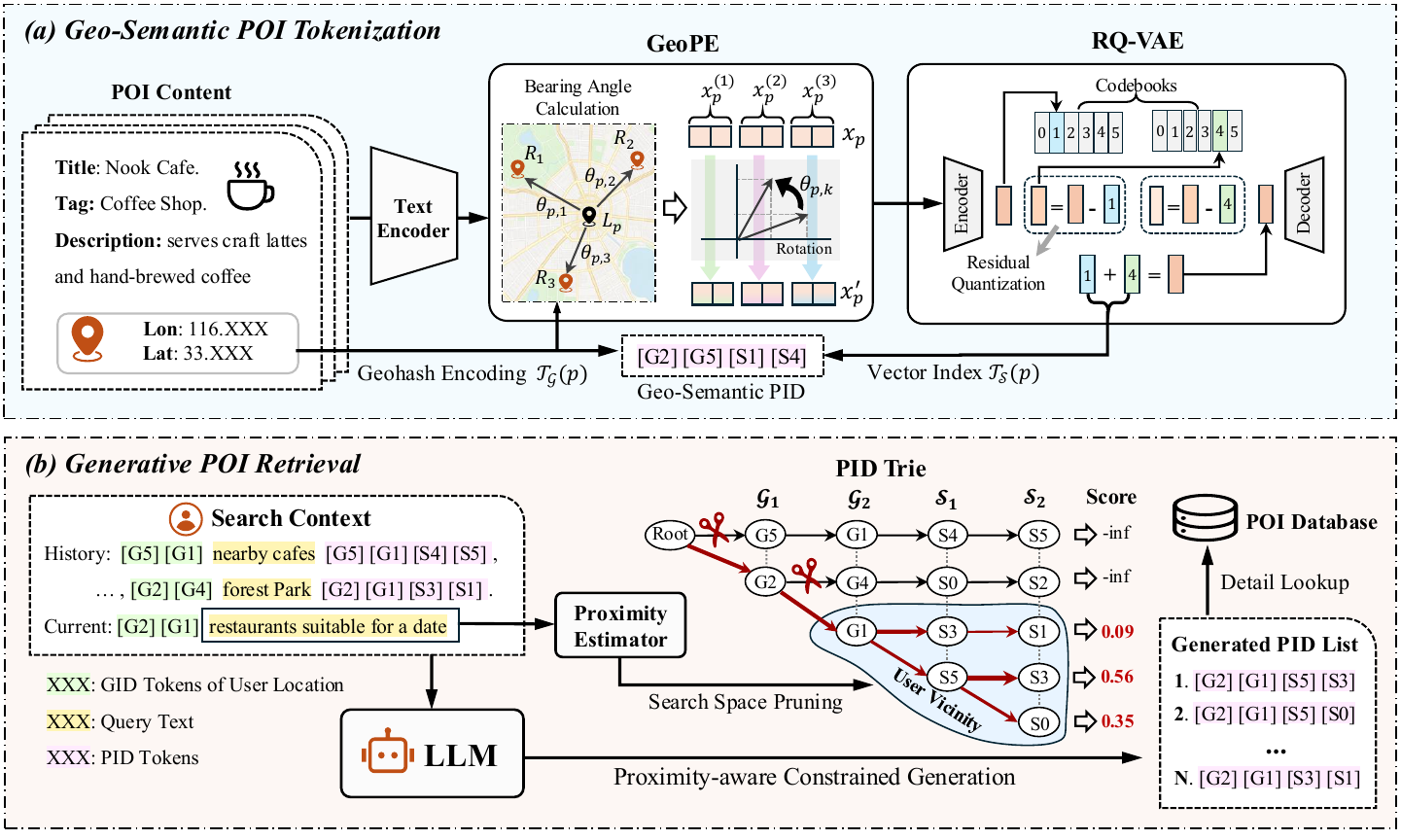}
  \caption{The overview of GenPOI framework. It consists of two modules: (a) Geo-Semantic POI Tokenization, which discretizes POIs into LLM-friendly token sequences; and (b) Generative POI Retrieval, where the LLM reasons over unified heterogeneous search contexts to generate relevant POIs.}
  \label{fw}
\end{figure*}

\subsection{Generative Retrieval}
With the rapid advancement of large language models (LLMs)~\cite{LLMSurvey1, LLMSurvey2}, Generative Retrieval (GR)~\cite{GRsurvey, GR-E-commerceRetrieval, OneSearch} has emerged as a promising paradigm, achieving notable performance in diverse domains including document retrieval~\cite{GR4Doc, GR4Doc-2} and recommender systems~\cite{LLM4Rec, LLM4POI, Onesug}. Distinct from traditional discriminative retrieval frameworks, GR reformulates the retrieval pipeline as an end-to-end autoregressive generation task. In the GR paradigm, each candidate entity is mapped to a discretized and semantically informative identifier, termed a semantic ID~\cite{sid1, sid2}. State-of-the-art semantic ID generation approaches mainly involve hierarchical clustering of latent embeddings~\cite{RQ-Kmeans} and residual vector quantization~\cite{RQ-VAE}, both designed to preserve the structural correlations among entities in a discrete representation space. By encoding the full corpus into the model's parameter space, LLMs learn the intrinsic semantic correlations among all SIDs during training. They are then able to autoregressively generate the exact semantic IDs for target entities by leveraging rich contextual information.

Although a few recent works~\cite{GNPR} begin to incorporate location into semantic IDs, they target next-POI recommendation rather than query-driven map search, leaving current generative retrieval paradigms still short of jointly grounding the LLM's vocabulary in geographic space and enforcing proximity-aware decoding for POI retrieval.

\section{Problem Formulation}
In this work, we focus on the POI retrieval task within general search scenarios. Formally, let $\mathcal{P} = \{p_1, p_2, \dots, p_N\}$ denote a large-scale repository of $N$ POIs. Each POI $p \in \mathcal{P}$ is represented by its geographic coordinates $L_p = (lat_p, lng_p) \in \mathbb{R}^2$ and a set of textual attributes $A_p$ (e.g., title, category, and other relevant descriptive metadata).
Given a user request, the available context is defined by the user's current location $L_u = (lat_u, lng_u)$, a textual search query $Q_u$, and a historical interaction sequence $H_u = \{(Q_{u,t}, L_{u,t}, p_{u,t})\}_{t=1}^T$, where $T$ denotes the sequence length. Here, $Q_{u,t}$ and $L_{u,t}$ represent the $t$-th query and the associated location, respectively, and $p_{u,t} \in \mathcal{P}$ denotes the specific POI clicked by the user.
The objective of POI retrieval is to learn a ranking function $f: (Q_u, L_u, H_u, \mathcal{P}) \rightarrow \mathcal{P}^K$ that maps the input context to a ranked list of $K$ relevant POIs:
\begin{equation}
P_u = f(Q_u, L_u, H_u, \mathcal{P}) = \{p_u^1, p_u^2, \dots, p_u^K\} \subseteq \mathcal{P}
\end{equation}
where $p_u^i$ denotes the $i$-th ranked POI in the result list. The core challenge of this task lies in effectively synergizing query semantics, spatial constraints, and historical behaviors to retrieve accurate and personalized POIs from an expansive search space.

\section{Methodology}

\subsection{Overview}
In this section, we introduce the proposed generative POI retrieval framework, \textbf{GenPOI}. Departing from conventional discriminative paradigms, GenPOI harmonizes the modeling of heterogeneous search contexts and reformulates the retrieval task as a sequence-to-sequence generation process. This framework effectively bridges the gap between existing generative retrieval paradigms and their real-world application in map search. As illustrated in Fig.~\ref{fw}, GenPOI comprises two core components:

\begin{itemize}
    \item \textbf{Geo-Semantic POI Tokenization:} This stage transforms heterogeneous POI attributes into discrete POI IDs (PIDs). Specifically, it integrates POI semantics with spatial features via a novel GeoPE module within the latent embedding space. These enriched embeddings are subsequently quantized into Semantic IDs (SIDs) and concatenated with explicit geohash-based Geographic IDs (GIDs) to yield the final PIDs, enabling LLMs to perceive discrete POIs within a unified geo-semantic vocabulary.
    \item \textbf{Generative POI Retrieval:} This stage leverages the powerful contextual modeling capabilities of LLMs to directly generate target PIDs conditioned on the sequential search context. To ensure both the validity and spatial plausibility of the results, we introduce proximity-aware constrained generation via a PID Token Trie, which is further adaptively pruned based on the user's location. This mechanism effectively narrows the generation space to efficiently produce high-quality retrieved POIs.
\end{itemize}

\subsection{Geo-Semantic POI Tokenization}
To distinguish the geographical nuances of POIs within a massive database, prevailing generative retrieval methods that rely solely on semantic IDs are insufficient, as they result in spatially agnostic learning by LLMs. To address this, we introduce Geo-Semantic POI Tokenization, which seamlessly integrates geographic context into the POI identifiers (PIDs) to facilitate LLM's spatial-aware POI perception.

\textbf{Explicit Geographic Identifiers.} For the unified location identifiers of POIs and users, we first incorporate explicit geographic identifiers (GIDs) using Geohash encoding~\cite{geohash}. This encoding mechanism bridges the gap between continuous numerical coordinates and the discrete vocabulary of LLMs. In detail, we leverage Geohash to discretize geographic coordinates into hierarchical spatial grids, mapping the continuous location of a POI $p$ into a discrete sequence of tokens. Formally, the GID is defined as:
\begin{equation}
\mathcal{T}_{\mathcal{G}}(L_p) = \text{Geohash}(\text{lat}_p, \text{lon}_p) = \{\mathcal{G}_1, \mathcal{G}_2, \dots, \mathcal{G}_{\ell_g}\}
\end{equation}
where $\ell_g$ denotes the precision level of the Geohash, and each $\mathcal{G}_i$ represents a specific token within the geographic vocabulary. A key advantage of this encoding scheme is its prefix-sharing property: POIs in close spatial proximity naturally share longer common prefixes in their GID sequences. Crucially, by applying this unified representation to both POI and user locations, we project them into a shared discrete space. This alignment enables the LLM to directly reason about the spatial interactions between users and POIs, effectively transforming geographical proximity into a recognizable token pattern during the generative process.

\textbf{Geographic Position Embedding.}
To complement the explicit GIDs with continuous spatial variance, we further propose a novel Geographic Position Embedding (GeoPE) that injects geographical semantics directly into the POI's representation. Initially, the textual attributes $A_{p}$ of a POI $p$ (i.e., its name and category) are concatenated into a structured template and fed into a pre-trained text encoder to obtain the base POI embedding $x_p \in \mathbb{R}^D$. To capture spatial dependencies, GeoPE implicitly encodes coordinates by rotating embeddings~\cite{rope} according to their relative orientations with respect to a set of reference points. Specifically, we establish a spatial coordinate system by identifying $\Omega$ reference points $\mathcal{R} = \{R_1, R_2, \dots, R_{\Omega}\}$, where each $R_{\omega} = (\text{lon}_{\omega}^{ref}, \text{lat}_{\omega}^{ref})$ is obtained via K-Means clustering on the POI distribution. These points serve as spatial anchors that represent regional density centers.

For each POI $p$ and reference point $\omega$, we compute the bearing angle $\theta_{p,\omega}$ to characterize their directional relationship:
\begin{equation}
\theta_{p,\omega} = \text{atan2}(\Delta \text{lat}_{p,\omega}, \Delta \text{lon}_{p,\omega} \cdot \cos(\text{lat}_{\omega}^{ref}))
\end{equation}
where $\Delta \text{lon}_{p,\omega} = \text{lon}_p - \text{lon}_{\omega}^{ref}$ and $\Delta \text{lat}_{p,\omega} = \text{lat}_p - \text{lat}_{\omega}^{ref}$. The angle $\theta_{p,\omega} \in [0, 2\pi)$ indicates the azimuth from anchor $\omega$ to POI $p$. Notably, the ensemble of $\Omega$ bearing angles uniquely determines the POI's 2D position through triangulation, providing comprehensive spatial specification. To implement this, we partition the embedding $x_p$ into $\Omega$ equal segments: $x_p = [x_p^{(1)}, x_p^{(2)}, \dots, x_p^{(\Omega)}]$, where each segment $x_p^{(\omega)} \in \mathbb{R}^{D/\Omega}$ corresponds to a specific reference point $\omega$. For each segment, we apply 2D rotation to consecutive dimension pairs governed by the bearing angle $\theta_{p,\omega}$:
\begin{equation}
\begin{bmatrix}
x_{p, 2j}^{(\omega)'} \\
x_{p, 2j+1}^{(\omega)'}
\end{bmatrix} =
\begin{bmatrix}
\cos \theta_{p,\omega} & -\sin \theta_{p,\omega} \\
\sin \theta_{p,\omega} & \cos \theta_{p,\omega}
\end{bmatrix}
\begin{bmatrix}
x_{p, 2j}^{(\omega)} \\
x_{p, 2j+1}^{(\omega)}
\end{bmatrix}
\end{equation}
for $j = 0, 1, \dots, D/(2\Omega)-1$. This transformation encodes the spatial relationship relative to reference point $\omega$ into the corresponding embedding subspace. The rotated segments are then concatenated to form the final rotated embedding:
\begin{equation}
x_p' = [x_p^{(1)'}, x_p^{(2)'}, \dots, x_p^{(\Omega)'}] \in \mathbb{R}^D.
\end{equation}
Through GeoPE, POIs at different locations receive distinct rotational transformations, ensuring that spatial context is implicitly preserved and distinguishable within the latent embedding space.

\textbf{Geo-Semantic PID Construction.} Following the GeoPE transformation, the geography-enriched POI embeddings are fed into a Residual Quantized Variational Autoencoder (RQ-VAE)~\cite{RQ-VAE, VQ-VAE} to derive semantic identifiers (SIDs). This architecture consists of an encoder network $\mathbf{E}_\phi$, a residual quantizer $\mathbf{Q}$, and a decoder network $\mathbf{D}_\psi$. Specifically, the encoder transforms the embedding $x_p' \in \mathbb{R}^D$ into a latent representation $\mathbf{h}_p = \mathbf{E}_\phi(x_p') \in \mathbb{R}^d$, where $d$ denotes the latent dimension. The residual quantizer discretizes $\mathbf{h}_p$ through a multi-stage process utilizing $\ell_s$ hierarchical codebooks $\{\mathcal{C}_1, \mathcal{C}_2, \dots, \mathcal{C}_{\ell_s}\}$, with each codebook $\mathcal{C}_\ell$ containing $M$ learnable code vectors. At each quantization stage $\ell \in \{1, \dots, \ell_s\}$, we identify the nearest code vector in $\mathcal{C}_\ell$ to quantize the residual from the preceding stage:
\begin{equation}
\mathbf{z}_{p,\ell} = \arg\min_{\mathbf{c} \in \mathcal{C}_\ell} \|\mathbf{r}_{p,\ell-1} - \mathbf{c}\|_2^2
\end{equation}
where $\mathbf{r}_{p,0} = \mathbf{h}_p$ and the residual for the subsequent stage is $\mathbf{r}_{p,\ell} = \mathbf{r}_{p,\ell-1} - \mathbf{z}_{p,\ell}$. The latent representation is reconstructed as $\hat{\mathbf{h}}_p = \sum_{\ell=1}^{\ell_s} \mathbf{z}_{p,\ell}$, allowing the decoder to recover the original embedding $\hat{x}_p' = D_\psi(\hat{\mathbf{h}}_p)$. The RQ-VAE is optimized by minimizing a joint reconstruction and commitment loss:
\begin{equation}
\mathcal{L}_{RQ} = \|x_p' - \hat{x}_p'\|_2^2 + \beta \sum_{\ell=1}^{\ell_s} \|\text{sg}[\mathbf{r}_{p,\ell-1}] - \mathbf{z}_{p,\ell}\|_2^2
\end{equation}
where $\text{sg}[\cdot]$ denotes the stop-gradient operation and $\beta$ is a balancing hyperparameter.

Since each quantized vector $\mathbf{z}_{p,\ell}$ corresponds to a discrete index $i_{p,\ell} \in \{0, 1, \dots, M-1\}$, these indices collectively form the semantic token sequence $\mathcal{T}_{\mathcal{S}}(p)=\{\mathcal{S}_1, \mathcal{S}_2, \dots, \mathcal{S}_{\ell_s}\}$ for each POI. Given that the input embeddings are spatially enriched via GeoPE, the resulting SIDs implicitly encapsulate both semantic and spatial nuances. Analogous to the prefix-sharing property of GIDs, the hierarchical nature of RQ-VAE ensures that POIs with high geo-semantic similarity tend to share common prefix tokens. Finally, we concatenate the explicit geographic tokens with these semantic tokens to constitute the complete PID: $\mathcal{T}(p) = \mathcal{T}_{\mathcal{G}}(L_p) \parallel \mathcal{T}_{\mathcal{S}}(p)$. To resolve potential identifier collisions where multiple POIs map to the same PID, we append a deduplication code to ensure unique identification. In practice, such collisions are rare as GIDs already provide high spatial discriminability.

\subsection{Generative POI Retrieval}

With geo-semantic PIDs constructed, we formulate POI retrieval as a generative sequence-to-sequence task.
Given a user's search context, which comprises historical records $H_u$ and the current query $Q_u$ and location $L_u$, our objective is to leverage LLMs to autoregressively generate the target PID token sequence $\mathcal{T}(p^*)$. To maintain geospatial consistency across the context sequence, all request locations are encoded using the same Geohash scheme as the POI tokenization, yielding a series of geographic tokens $\{\mathcal{T}_{\mathcal{G}}(L_{u,1}), \dots, \mathcal{T}_{\mathcal{G}}(L_{u,T}), \mathcal{T}_{\mathcal{G}}(L_u)\}$. The complete input sequence is constructed as follows:
\begin{equation}
\begin{aligned}
&\mathcal{S}_{input} = [\mathcal{T}_{\mathcal{G}}(L_{u,1}) \parallel \mathcal{T}_{text}(Q_{u,1}) \parallel \mathcal{T}(p_{u,1})  \cdots \\
&\mathcal{T}_{\mathcal{G}}(L_{u,T}) \parallel \mathcal{T}_{text}(Q_{u,T}) \parallel \mathcal{T}(p_{u,T}) \parallel \mathcal{T}_{\mathcal{G}}(L_u) \parallel \mathcal{T}_{text}(Q_u)],
\end{aligned}
\end{equation}
where $\mathcal{T}_{\text{text}}(\cdot)$ denotes the tokenized query text and $\mathcal{T}(p_{u,t}) = \mathcal{T}_{\mathcal{G}}(p_{u,t}) \parallel \mathcal{T}_{\mathcal{S}}(p_{u,t})$ represents the PID of the historical interacted POI at time step $t$. The LLM is trained to predict the PID for target POIs by maximizing the autoregressive likelihood over the PID token space, formalized as follows:
\begin{equation}
P(\mathcal{T}(p^*) \mid \mathcal{S}_{input}) = \prod_{i=1}^{|\mathcal{T}(p^*)|} P(v_i^* \mid v_{<i}^*, \mathcal{S}_{input}).
\end{equation}
In this formulation, $v_i^*$ denotes the $i$-th token of the target sequence $\mathcal{T}(p^*)$, and $v_{<i}^*$ represents the prefix generated prior to step $i$. To obtain a ranked list of candidates, we employ beam search~\cite{beamsearch} during the inference phase. This decoding strategy explores multiple high-probability paths in the PID token space to retrieve the top-$K$ candidate POIs $P_u$ that best satisfy the user's latent intent and geographical constraints.

\textbf{PID Trie Constrained Generation.}
Direct token generation from the entire LLM vocabulary often yields invalid PIDs that do not correspond to any POI in the database. To ensure the structural integrity and validity of the generated sequences, we construct a prefix tree (Trie)~\cite{cgen} encompassing the complete set of PID sequences $\{\mathcal{T}(p)\}_{p \in \mathcal{P}}$ within the database $\mathcal{P}$. During each decoding step $i$, the model's output distribution is constrained by a mask mechanism that only permits tokens forming valid prefixes according to the Trie:
\begin{equation}
P(v_i^* \mid v_{<i}^*, \mathcal{S}_{input}) = \frac{\exp(\sigma(v_i^*) / \tau) \cdot \mathbb{I}[v_i^* \in \mathcal{V}_i]}{\sum_{v' \in \mathcal{V}_i} \exp(\sigma(v') / \tau)}
\end{equation}
where $\sigma(\cdot)$ denotes the logit produced by the LLM, $\tau$ is the temperature parameter, and $\mathcal{V}_i$ represents the set of permissible tokens at position $i$ conditioned on the previously generated prefix $v_{<i}^*$. Concretely, the permissible token set $\mathcal{V}_i$ at decoding step $i$ is determined by traversing the PID Trie. Given the already-generated prefix $v_{<i}^* = (v_1^*, \dots, v_{i-1}^*)$, we follow the corresponding path in the Trie and collect all children of the current node as $\mathcal{V}_i$:
\begin{equation}
\mathcal{V}_i = \text{Children}\big(\text{Trie}[v_1^*, v_2^*, \dots, v_{i-1}^*]\big)
\end{equation}
This ensures that at each step, only tokens leading to valid (existing) PID sequences are permitted. By integrating beam search with this constrained vocabulary, the framework guarantees that every generated PID maps to a physically existent POI in our database. Notably, when new POIs are added, their PIDs can be obtained via the frozen tokenization pipeline and inserted into the Trie incrementally, avoiding LLM retraining for typical updates.

\textbf{Proximity-Aware Search Space Pruning.}
Despite the application of constrained decoding, the inherent stochasticity of autoregressive generation and the massive scale of the POI database may still cause the LLM to generate spatially irrelevant results that are far from the user's current location. This module is motivated by the observation that map searches are predominantly localized, while distinct queries exhibit varying spatial proximity requirements (as shown in Fig.~\ref{query}). Specifically, we leverage a lightweight proximity estimator, a classifier trained on query-POI pairs, to adaptively predict the geographic proximity level $\lambda(Q_u) \in \{0, 1, \dots, \ell_{\mathcal{G}}\}$ of a user query $Q_u$. A higher value of $\lambda(Q_u)$ indicates a more localized search intent, necessitating a more restricted search space.
Once $\lambda(Q_u)$ is determined, we enforce a hard constraint that the prefix of the target PID must align with the user's current location $L_u$:
\begin{equation}
\mathcal{T}_{\mathcal{G}}^{(1:\lambda(Q_u)-\gamma)}(p^*) = \mathcal{T}_{\mathcal{G}}^{(1:\lambda(Q_u)-\gamma)}(L_u)
\end{equation}
where $(1{:}\lambda(Q_u){-}\gamma)$ denotes the first $\lambda(Q_u){-}\gamma$ Geohash tokens of the GID, and $\gamma \in \mathbb{Z}^+$ denotes a relaxation offset that prevents over-restriction of the search space and accounts for potential boundary effects. By fixing the initial geographic tokens, this approach effectively prunes the search space by leveraging the spatial hierarchical nature of the GID. Consequently, it inherently prevents the hallucination of distant POIs, ensuring spatially pertinent retrieval. Meanwhile, this mechanism accelerates LLM inference by reducing the number of active decoding steps during beam search, as the initial geographic tokens are pre-filled based on the user's location rather than autoregressively decoded.

\section{Experiment Results and Analysis}

\subsection{Experimental Setting}

\subsubsection{\textbf{Datasets}} We evaluate the proposed GenPOI using two real-world general map search datasets from Tencent Maps. In detail, (1) \textbf{TMap-S} (Small-scale) focuses on POIs from 5 representative cities in China, providing a controlled environment to evaluate retrieval accuracy in dense urban areas. (2) \textbf{TMap-L} (Large-scale) spans a nationwide geographic scope with over 10 million POIs, designed to assess the scalability and robustness of GenPOI in extensive search scenarios. For the user search data, we collect 5 months of online general search logs from the map service and organize them into user-level sequences. Each sample represents a sequence of historical interaction records, where each interaction captures the association among a user query, its geospatial context, and the finally clicked POI. Specifically, the ground-truth POI of each query is defined as the one finally clicked by the user with sufficient dwell time, which serves as a reliable proxy of the user's true intent. Detailed statistics of the datasets are summarized in Table~\ref{dataset}.

\begin{table}[ht]
\centering
\caption{Statistics of TMap-S and TMap-L datasets.}
\label{dataset}
\begin{tabular*}{0.7\columnwidth}{@{}l@{\extracolsep{\fill}}cc@{}}
\toprule
\textbf{Metric} & \textbf{TMap-S} & \textbf{TMap-L} \\ \midrule
\# POIs & 1.25M & 11.62M \\
\# User Sequences & 1.09M & 4.77M \\
\# Interactions & 3.50M & 17.61M \\
Avg. History Length & 3.20 & 3.69 \\
Geographic Coverage & 5 Cities & Nationwide \\ \bottomrule
\end{tabular*}
\end{table}

\begin{table*}[htbp]
  \centering
  \caption{Comparison of POI Retrieval Performance of Different Methods. The best and second-best results are highlighted in bold and underline, respectively.}
  \label{tab:poi_retrieval_performance}
  \begin{tabular*}{\textwidth}{@{}l@{\extracolsep{\fill}}cccccccccccc@{}}
    \toprule
    \multirow{3}{*}{Methods} & \multicolumn{6}{c}{\textbf{TMap-S}} & \multicolumn{6}{c}{\textbf{TMap-L}} \\
    \cmidrule(lr){2-7} \cmidrule(lr){8-13}
    & Recall & NDCG & Recall & NDCG & Recall & NDCG & Recall & NDCG & Recall & NDCG & Recall & NDCG \\
    & @5 & @5 & @10 & @10 & @20 & @20 & @5 & @5 & @10 & @10 & @20 & @20 \\
    \midrule
    DSSM & 0.2018 & 0.1503 & 0.2632 & 0.1702 & 0.3352 & 0.1892 & 0.1130 & 0.0813 & 0.1520 & 0.0967 & 0.1901 & 0.1113 \\
    DSSM-dist & 0.3510 & 0.2825 & 0.4109 & 0.3018 & 0.4787 & 0.3189 & 0.3392 & 0.2598 & 0.4041 & 0.2809 & 0.4582 & 0.2946 \\
    PALM & 0.3343 & 0.2619 & 0.4115 & 0.2871 & 0.5055 & 0.3108 & 0.2930 & 0.1735 & 0.3548 & 0.1934 & 0.4170 & 0.2092 \\
    HGAMN & 0.4042 & 0.3524 & 0.4716 & 0.3945 & 0.5440 & 0.4005 & 0.3573 & 0.2905 & 0.4020 & 0.3086 & 0.4310 & 0.3161 \\
    TIGER-0.6B & 0.5312 & 0.4447 & 0.5707 & 0.4576 & 0.5923 & 0.4632 & 0.4497 & 0.3869 & 0.4857 & 0.3986 & 0.5117 & 0.4052 \\
    TIGER-1.7B & 0.5549 & 0.4678 & 0.5889 & 0.4789 & 0.6093 & 0.4842 & 0.4987 & 0.4262 & 0.5333 & 0.4375 & 0.5530 & 0.4425\\
    GNPR-SID-0.6B & 0.5385 & 0.4500 & 0.5838 & 0.4649 & 0.6072 & 0.4708 & 0.4567 & 0.3962 & 0.4933 & 0.4088 & 0.5210 & 0.4160 \\
    GNPR-SID-1.7B & \underline{0.5594} & \underline{0.4717} & \underline{0.5959} & \underline{0.4836} & \underline{0.6184} & \underline{0.4895} & \underline{0.5085} & \underline{0.4367} & \underline{0.5481} & \underline{0.4492} & \underline{0.5640} & \underline{0.4538} \\
    \midrule
    \textbf{GenPOI-0.6B} & 0.6625 & 0.5593 & 0.7087 & 0.5745 & 0.7340 & 0.5809 & 0.5967 & 0.5033 & 0.6357 & 0.5160 & 0.6553 & 0.5210 \\
    \textbf{GenPOI-1.7B} & \textbf{0.6820} & \textbf{0.5758} & \textbf{0.7294} & \textbf{0.5913} & \textbf{0.7568} & \textbf{0.5983} & \textbf{0.6193} & \textbf{0.5280} & \textbf{0.6577} & \textbf{0.5407} & \textbf{0.6863} & \textbf{0.5478} \\
    \bottomrule
  \end{tabular*}
\end{table*}

\subsubsection{\textbf{Evaluation Metric \& Comparison Method}}  To assess the performance of GenPOI, we employ standard retrieval metrics including Recall@K and Normalized Discounted Cumulative Gain (NDCG@K) with K values of 5, 10, and 20. Specifically, Recall@K measures the success rate of retrieving the ground-truth within the top-K results, while NDCG@K evaluates ranking quality by penalizing lower positions of the target POI.

We compare GenPOI against a wide range of state-of-the-art baselines to demonstrate its effectiveness. \textbf{(1) Neural embedding-based methods}: we include DSSM~\cite{DSSM} as a classic semantic matching model, along with its variant DSSM-dist, which is tailored for map scenarios through a distance-based re-ranking stage. Furthermore, we adopt specialized geospatial models including PALM~\cite{PALM} and HGAMN~\cite{hgamn}, which leverage an attention mechanism and heterogeneous graph attention networks to incorporate geospatial considerations into the retrieval pipeline. \textbf{(2) Generative retrieval methods}: we consider TIGER~\cite{tiger} and GNPR-SID~\cite{GNPR} as representative baselines typically used for item/POI recommendation tasks. To ensure a fair comparison, both generative baselines are adapted to our map search scenario by integrating Geohash codes of the user's location and query text into the input context of the LLM. Additionally, textual POI addresses are incorporated during semantic ID quantization.

\subsubsection{\textbf{Implementation Detail}} Regarding the architecture of GenPOI, we set the Geohash precision (i.e., GID token length) to $\ell_g=6$ and the number of reference points in GeoPE to $\Omega=16$. All tokens associated with semantic IDs and geographic IDs are integrated into the LLM as special tokens within an extended vocabulary, allowing them to be learned from scratch during the training process. The RQ-VAE component is configured with $\ell_s=3$ hierarchical codebooks, each containing $M=128$ latent vectors. For text encoding, we utilize the pre-trained BGE-M3~\cite{bge-m3} as the text encoder. The Proximity Estimator also adopts BGE-M3 as its backbone, where only an additional classification head is trained to ensure efficient learning. For both GenPOI and generative baselines, we employ Qwen3-0.6B and Qwen3-1.7B~\cite{qwen3} as the LLM backbones, performing full-parameter supervised fine-tuning. 

All experiments are conducted on $4$ NVIDIA A100 GPUs. Our model is implemented using the PyTorch~\cite{paszke2019pytorch} framework. For the LLM training, we employ a per-device training batch size of 256 with gradient accumulation steps set to 4. The model is optimized for 5 epochs using a learning rate of $1.0 \times 10^{-4}$. During the inference phase, we use beam search where the beam size is set to the number of target POIs $K$, and the relaxation factor $\gamma$ is set to 2. For the RQ-VAE, the hidden dimensions are set to $[512, 256, 128]$ and the code vector embedding dimension is set to 32. This component is trained with a commitment loss weight $\beta$ of 0.25 and a learning rate of $5.0 \times 10^{-4}$. To ensure practical relevance and data quality, we curate our POI collection by ranking POIs based on their historical click-through rates and selecting the most frequent entities from a large-scale candidate pool on Tencent Map.

\subsection{Model Performance}
Table \ref{tab:poi_retrieval_performance} presents the comprehensive evaluation results of POI retrieval performance on two datasets, TMap-S and TMap-L. Particularly, we derive the following key observations:

\noindent\textbf{(1) Our proposed GenPOI consistently achieves superior performance across all evaluation metrics and datasets.} It achieves a maximum improvement of 13.84\% in Recall@20 compared to the strongest baseline. This significant gain is attributed to the generative retrieval paradigm tailored for geospatial contexts. Unlike traditional discriminative frameworks, GenPOI facilitates unified context modeling, enabling LLM-based sophisticated intent reasoning to resolve underspecified and complex queries. Furthermore, GenPOI introduces specialized geospatial modeling considerations to enhance the task-oriented adaptability, thereby fully unleashing the powerful context-modeling capacity of LLMs in capturing the intricate spatial-semantic dependencies inherent in POI retrieval.

\noindent\textbf{(2) Traditional neural embedding-based methods yield suboptimal performance compared to generative methods.} These models primarily rely on dual-tower architectures for coarse-grained semantic matching, which lack the discriminative power required for massive POI databases and general search queries. Moreover, these methods tend to process heterogeneous search contexts independently, making it difficult to achieve the comprehensive intent reasoning required for context-dependent query understanding.

\noindent\textbf{(3) Existing generative retrieval methods exhibit poor adaptability to map search scenarios.} As shown in Table~\ref{tab:poi_retrieval_performance}, while baselines like TIGER and GNPR-SID outperform embedding-based models, they still fall significantly short of GenPOI. This gap arises primarily because these methods lack an effective and unified modeling of geospatial factors, making it difficult to achieve a profound spatial understanding of both user search contexts and POIs. Consequently, they often fail to capture critical proximity constraints within the expansive generation space, leading to degraded retrieval precision.

\begin{table}[t]
  \centering
  \caption{Ablation study setup. EGI, GeoPE, TCG, and SSP denote explicit geographic identifiers for POIs, geographic position embedding, PID trie constrained generation, and proximity-aware search space pruning, respectively.}
  \label{tab:abla}
  \renewcommand{\arraystretch}{1.15}
  \begin{tabular*}{\columnwidth}{@{\extracolsep{\fill}}lcccc@{}}
    \toprule
    Model & EGI & GeoPE & TCG & SSP \\
    \midrule
    GenPOI$_{a-}$ & $\times$ & \checkmark & \checkmark & $-$ \\
    GenPOI$_{b-}$ & $\times$ & $\times$ & \checkmark & $-$ \\
    GenPOI$_{c-}$ & \checkmark & $\times$ & \checkmark & \checkmark \\
    GenPOI$_{d-}$ & \checkmark & \checkmark & $\times$ & \checkmark \\
    GenPOI$_{e-}$ & \checkmark & \checkmark & \checkmark & $\times$ \\
    GenPOI$_{f-}$ & \checkmark & \checkmark & $\times$ & $\times$ \\
    \midrule
    \textbf{GenPOI} & \checkmark & \checkmark & \checkmark & \checkmark \\
    \bottomrule
  \end{tabular*}
  \vspace{2pt}
  \begin{flushleft}
    Note: ``$-$'' indicates that SSP is not applicable as it necessitates EGI's spatial hierarchy.
  \end{flushleft}
\end{table}

\subsection{Analysis of Model Design}

\subsubsection{\textbf{Ablation Study}} To evaluate the contributions of the different modules of our model to the POI retrieval performance, we conduct an ablation study by comparing six variants: GenPOI$_{a-}$, GenPOI$_{b-}$, GenPOI$_{c-}$, GenPOI$_{d-}$, GenPOI$_{e-}$, and GenPOI$_{f-}$. The specific setting is outlined in Table~\ref{tab:abla}, where different models comprise various combinations of four modules.

\begin{figure}[ht]
  \centering
  \includegraphics[width=\columnwidth]{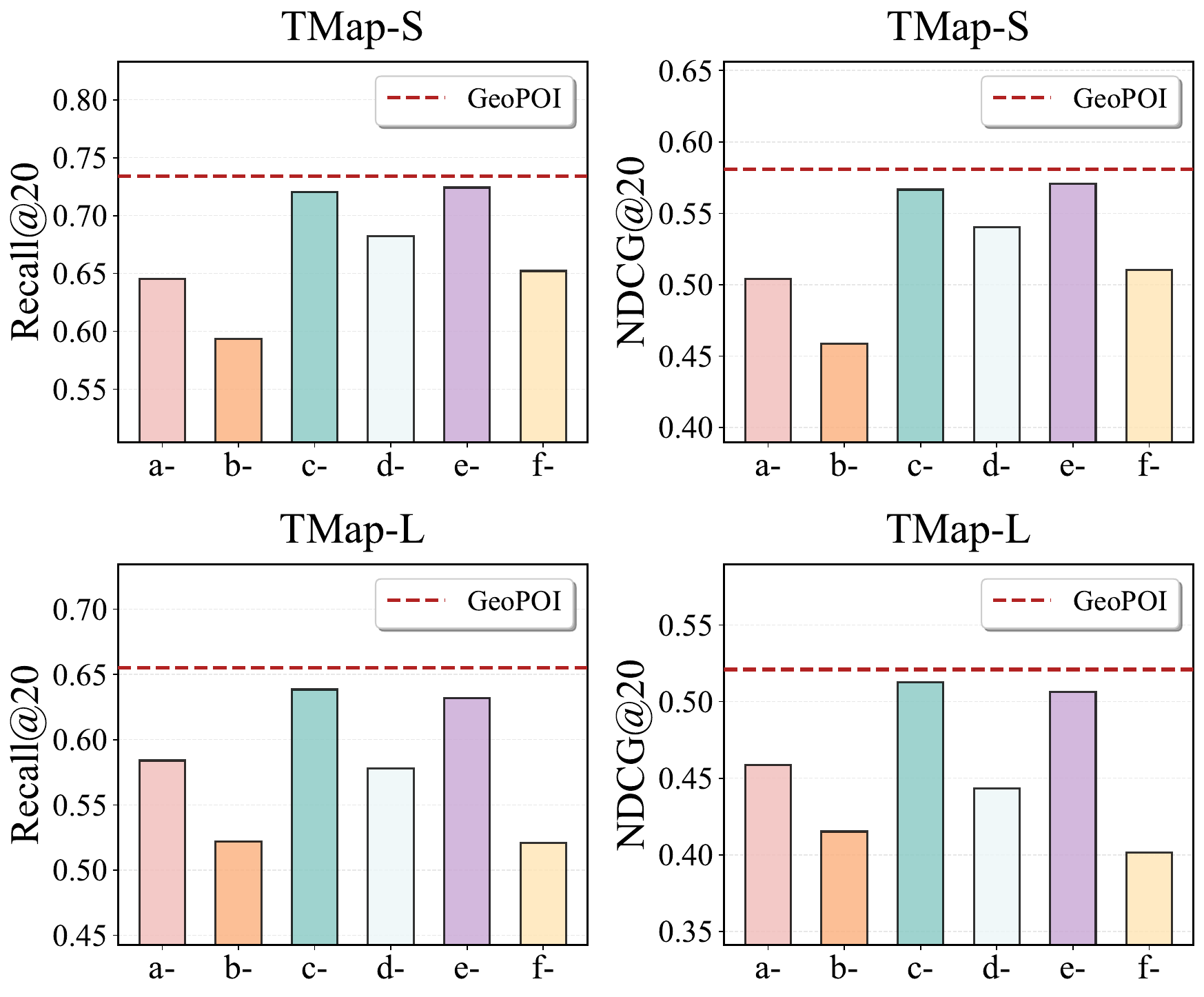}
  \caption{Results of ablation study. All variants are evaluated based on GenPOI-0.6B.}
  \label{abla}
\end{figure}

As shown in Fig.~\ref{abla}, the performance disparity between GenPOI and its variants highlights the necessity of each component. Specifically, GenPOI$_{a-}$ and GenPOI$_{b-}$ exhibit a significant performance decline, demonstrating that explicit geographic identifiers are critical for the LLM to understand the spatial relationship between users and POIs. The fact that GenPOI$_{a-}$ outperforms GenPOI$_{b-}$ suggests that GeoPE injects spatial semantics into the latent space, facilitating the generation of more geospatially discriminative SIDs. Furthermore, the results for GenPOI$_{d-}$, GenPOI$_{e-}$, and GenPOI$_{f-}$ underscore the importance of structural constraints, where TCG ensures the validity of the generated PID sequences and SSP enhances retrieval accuracy by narrowing down the candidate space from a spatial perspective. Notably, GenPOI$_{f-}$ exhibits the most pronounced performance drop on the TMap-L dataset, confirming that the absence of generation constraints leads to substantial degradation in large-scale industrial scenarios.

\subsubsection{\textbf{Spatial Variance Injection via GeoPE}} To intuitively demonstrate the impact of GeoPE on the POI embedding space, we employ t-SNE~\cite{TSNE} to visualize the POI embeddings before and after applying the GeoPE. In the initial distribution, embeddings generated by the text encoder are predominantly clustered by their semantic categories, where POIs belonging to the same category are tightly grouped regardless of their actual locations. Upon applying GeoPE, the embedding distribution undergoes a significant structural refinement. While the broad semantic clusters are preserved, the individual POI embeddings within each category become further stratified and dispersed based on their spatial positions. This shift demonstrates that GeoPE effectively injects spatial variance into the embedding level, transforming discrete geographical coordinates into continuous vector-level distinctions, thereby allowing the subsequent RQ-VAE to capture more granular POI differences.

\subsubsection{\textbf{Acceleration and Spatial Consistency via SSP}} Table~\ref{tab:ssp_impact} compares GenPOI variants to evaluate the impact of Proximity-Aware Search Space Pruning (SSP). The integration of SSP yields gains in both retrieval efficiency and spatial reliability. Specifically, SSP reduces inference latency by employing a lightweight proximity estimator to adaptively bypass irrelevant geographic tokens conditioned on the query. This optimization enables the LLM to skip the initial token generation phases, thereby avoiding the computational overhead of expansive beam searches across unrelated regions of the global POI space. Furthermore, SSP significantly lowers the spatial outlier rate by enforcing hard constraints centered on the user's vicinity. This mechanism effectively mitigates ``spatial hallucinations'', in which the model would otherwise generate semantically related but geographically distant results, thereby ensuring high spatial consistency in real-world deployment.

\begin{figure}[t]
  \centering
  \includegraphics[width=0.85\columnwidth]{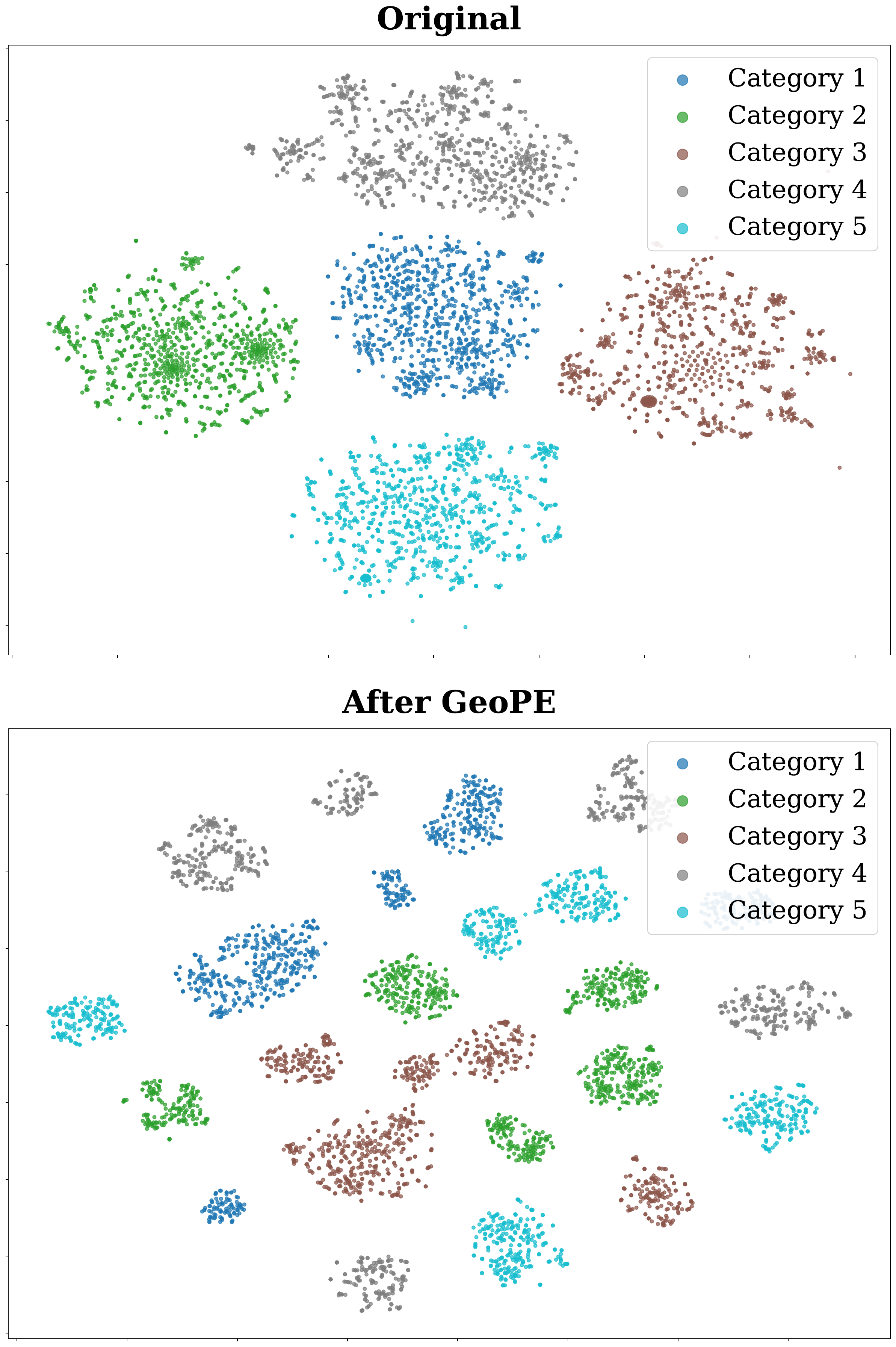}
  \caption{T-SNE visualization of POI embeddings before and after GeoPE encoding.}
  \label{emb}
\end{figure}

\begin{table}[t]
  \centering
  \caption{Impact of SSP on retrieval efficiency and spatial consistency. Spatial Outlier Rate indicates the ratio of retrieved POIs with a distance to user $>10\times$ that of the ground-truth.}
  \label{tab:ssp_impact}
  \resizebox{\columnwidth}{!}{%
  \begin{tabular}{l cc cc}
    \toprule
    \multirow{2}{*}{\textbf{Model Setting}} & \multicolumn{2}{c}{\textbf{Retrieval Time (ms)}} & \multicolumn{2}{c}{\textbf{Spatial Outlier Rate (\%)}} \\
    \cmidrule(lr){2-3} \cmidrule(lr){4-5}
    & \textbf{$K=10$} & \textbf{$K=20$} & \textbf{$K=10$} & \textbf{$K=20$} \\
    \midrule
    GenPOI-0.6B (w/o SSP) & 56.3 & 106.2 & 5.69 & 5.45 \\
    GenPOI-0.6B & 45.6 & 88.1 & 0.93 & 0.90 \\
    \textit{Speedup / Variation} & \textit{1.23$\times$} & \textit{1.20$\times$} & \textit{-4.76} & \textit{-4.55} \\
    \midrule
    GenPOI-1.7B (w/o SSP) & 76.5 & 144.3 & 4.78 & 4.59 \\
    GenPOI-1.7B & 62.4 & 122.0 & 0.59 & 0.54 \\
    \textit{Speedup / Variation} & \textit{1.23$\times$} & \textit{1.18$\times$} & \textit{-4.19} & \textit{-4.05} \\
    \bottomrule
  \end{tabular}%
  }
\end{table}

\begin{figure*}[ht]
  \centering
  \includegraphics[width=\textwidth]{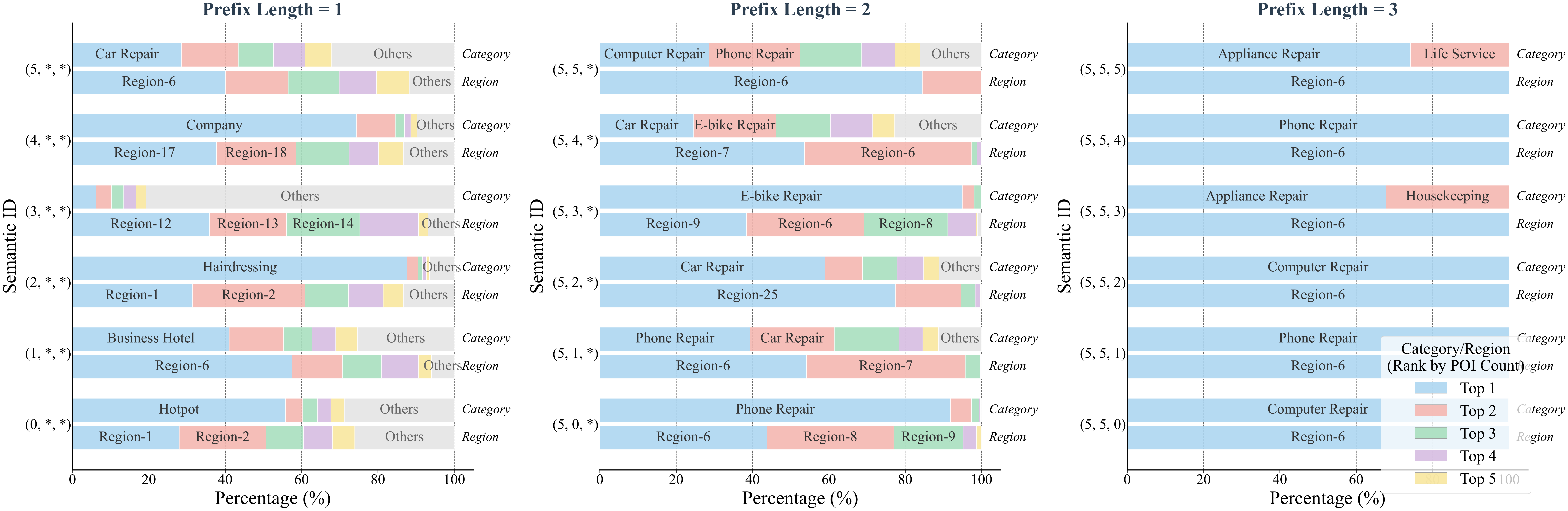}
  \caption{Distribution of categories and regions across semantic ID hierarchies. The Prefix Length denotes the number of fixed initial tokens used to aggregate POIs sharing the same semantic prefix.}
  \label{sid}
\end{figure*}

\begin{table}[t]
  \centering
  \caption{Impact of GID Length on POI Retrieval Performance.}
  \label{tab:gid_length_impact}
  \resizebox{\columnwidth}{!}{%
  \begin{tabular}{cc cccc}
    \toprule
    \textbf{GID} & \textbf{Geographic} & \textbf{Recall} & \textbf{NDCG} & \textbf{Recall} & \textbf{NDCG} \\
    \textbf{Length} & \textbf{Precision} & \textbf{@10} & \textbf{@10} & \textbf{@20} & \textbf{@20} \\
    \midrule
    3 & $\pm$ 78 km & 0.5537 & 0.4328 & 0.5856 & 0.4409 \\
    4 & $\pm$ 20 km & 0.6066 & 0.4742 & 0.6455 & 0.4841 \\
    5 & $\pm$ 2.4 km & 0.7033 & 0.5523 & \textbf{0.7403} & 0.5606 \\
    6 & $\pm$ 610 m & \textbf{0.7087} & \textbf{0.5745} & 0.7340 & \textbf{0.5809} \\
    7 & $\pm$ 76 m & 0.6561 & 0.5426 & 0.6649 & 0.5449 \\
    8 & $\pm$ 19 m & 0.6087 & 0.5076 & 0.6157 & 0.5094 \\
    \bottomrule
  \end{tabular}%
  }
\end{table}

\subsection{Impact of GID Length on Spatial Reasoning}
The length of geospatial IDs (GID) is a pivotal factor in the model's capacity to perceive and reason about spatial relationships. As illustrated in Table~\ref{tab:gid_length_impact}, retrieval performance improves consistently as the GID Length increases from 3 to 6 ($\pm$78 km to $\pm$610 m). This enhancement stems from the fact that finer spatial granularity enables the LLM to model the geometric alignment between user queries and POI locations with higher fidelity. However, performance begins to plateau or deteriorate beyond a length of 6. We attribute this to the fact that excessively high precision introduces a representational bottleneck: the resulting fragmentation of the search space hinders the model's ability to learn robust spatial patterns. Furthermore, longer GIDs increase the token count per identifier, imposing overhead on decoding complexity and inference latency. Therefore, a GID length of 6 offers sufficient geographic precision for effective POI retrieval.

\subsection{Hierarchical Semantic of SID}
As illustrated in Fig.~\ref{sid}, the distribution of Semantic IDs (SID) reveals a profound coarse-to-fine clustering effect that effectively encodes multi-level semantic features of POIs. At shallower prefix levels (Prefix Length = 1), clusters correspond to broad POI categories such as ``Life Service'' or ``Car Repair''. As the hierarchy deepens to Prefix Length 2 and 3, the SID transcends these macro-categorizations to characterize nuanced intra-category distinctions. This is evidenced by the alignment of specific third-level identifiers with fine-grained functional tags such as ``Phone Repair'' or ``Computer Maintenance''. With the integration of GeoPE, this hierarchical characteristic extends into the geo-semantic dimension. At deeper levels, SIDs exhibit a strong tendency to align with specific geographic regions. This progressive refinement indicates that the SID does not merely group POIs by surface-level categories but also captures their deep semantic and geographic nuances.

\subsection{Impact of Historical Context}
To investigate the influence of user behavioral history, we conduct a comparative analysis between GenPOI variants with and without historical context. As illustrated in Table~\ref{tab:main_results}, incorporating historical trajectories generally yields performance improvements across all metrics. This gain can be attributed to the model's ability to capture personalized user preferences, which provide critical context for disambiguating search intents. However, it is noteworthy that the performance gap between the two settings is not overly substantial. This observation suggests that our GenPOI framework possesses a robust capability to understand and retrieve POIs primarily based on the current query and geographic signals. Such a characteristic is particularly advantageous in real-world search scenarios, as it indicates the model's potential to effectively mitigate the cold-start problem for new users or those without historical data.

\begin{table}[htbp]
  \centering
  \caption{Performance comparison of GenPOI variants across different model scales, with and without historical context.}
  \label{tab:main_results}
  \resizebox{\columnwidth}{!}{%
    \begin{tabular}{lcccccc}
    \toprule
    Dataset & Model & History & Recall & NDCG & Recall & NDCG \\
            &       &         & @10    & @10  & @20    & @20  \\
    \midrule
    \multirow{4}{*}{TMap-S} & \multirow{2}{*}{GenPOI-0.6B} & $\times$ & 0.6777 & 0.5398 & 0.7084 & 0.5478 \\
          &       & \checkmark & 0.7087 & 0.5745 & 0.7340 & 0.5809 \\
\cmidrule{2-7}          & \multirow{2}{*}{GenPOI-1.7B} & $\times$ & 0.6992 & 0.5507 & 0.7270 & 0.5578 \\
          &       & \checkmark & 0.7294 & 0.5913 & 0.7568 & 0.5983 \\
    \midrule
    \multirow{4}{*}{TMap-L} & \multirow{2}{*}{GenPOI-0.6B} & $\times$ & 0.5803 & 0.4673 & 0.6023 & 0.4729 \\
          &       & \checkmark & 0.6357 & 0.5160 & 0.6553 & 0.5210 \\
\cmidrule{2-7}          & \multirow{2}{*}{GenPOI-1.7B} & $\times$ & 0.6120 & 0.4943 & 0.6390 & 0.5012 \\
          &       & \checkmark & 0.6577 & 0.5280 & 0.6863 & 0.5478 \\
    \bottomrule
    \end{tabular}%
  }
\end{table}

\subsection{Spatial Proximity Analysis of Queries}
\begin{figure}[t]
  \centering
  \includegraphics[width=\columnwidth]{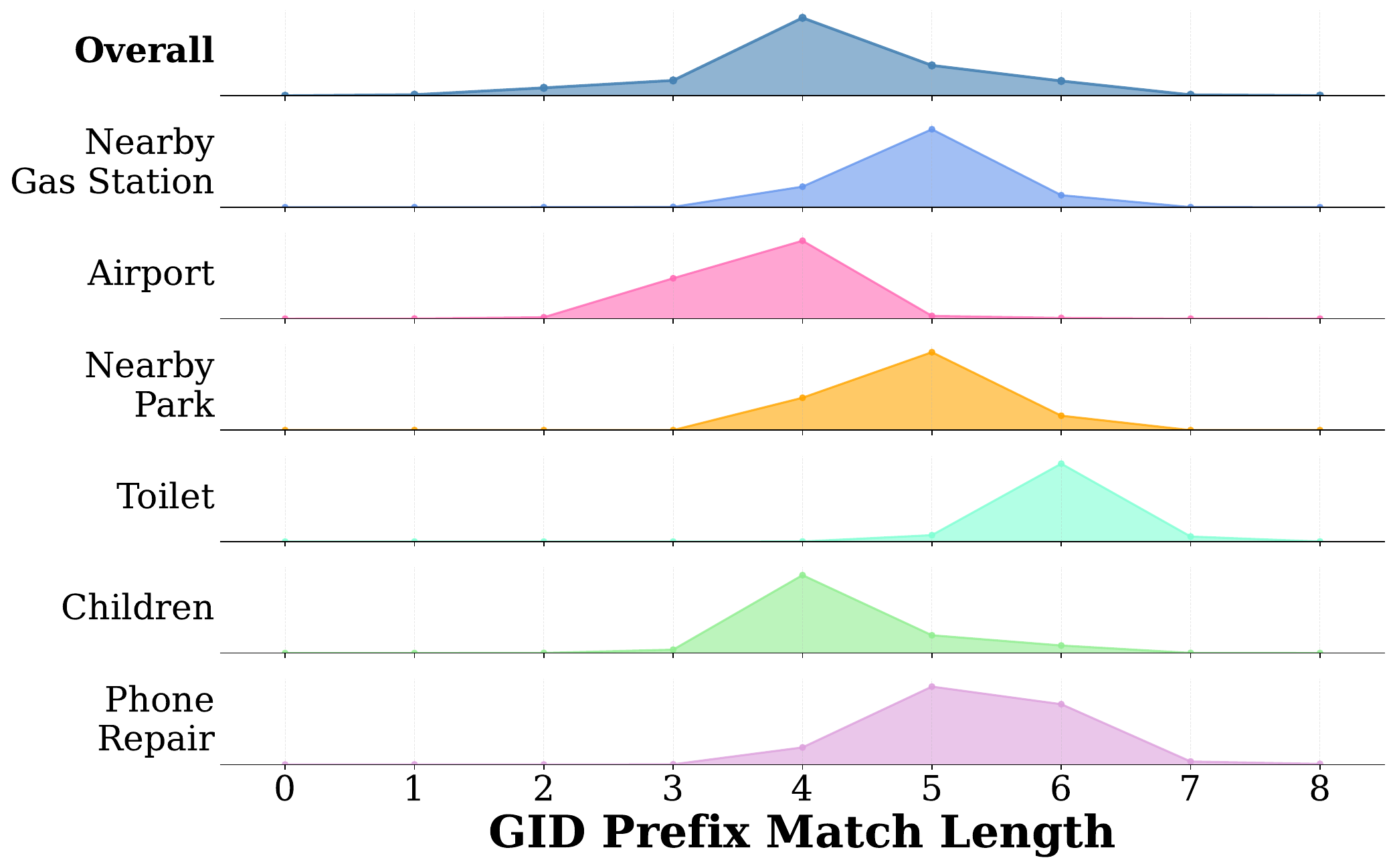}
  \caption{The distribution of maximum GID prefix match lengths between the user location and the clicked POI location for the overall dataset (TMap-L) and various high-frequency queries.}
  \label{query}
\end{figure}
The distribution of maximum GID prefix match lengths between user locations and the clicked POI locations in Fig.~\ref{query} reveals that general map search in most cases is inherently a localized task, with the majority of queries, including the overall distribution and various high-frequency categories, concentrated at a match length of 4 or higher. Furthermore, the distinct peak variations observed across different intents, such as the sharp localization of ``Toilet'' at length 6 compared to the broader regional intent of ``Airport'' at length 4, validate the motivation for our proximity estimator to adaptively predict $\lambda(Q_u)$. This adaptive mechanism allows the model to effectively prune the search space by leveraging the spatial hierarchical nature of the GID, skipping redundant geographic tokens for hyper-local queries while maintaining the necessary breadth for distant intents.

\subsection{Validity Analysis of Generated PIDs}
\begin{table}[htbp]
  \centering
  \caption{Invalid Generation Rate (\%) of GenPOI (0.6B) variants on TMap-S and TMap-L datasets. TCG indicates PID trie constrained generation.}
  \label{tab:invalid_rate}
  \begin{tabular}{lcccccc}
    \toprule
    \multirow{2}{*}{Model} & \multicolumn{3}{c}{TMap-S} & \multicolumn{3}{c}{TMap-L} \\
    \cmidrule(lr){2-4} \cmidrule(lr){5-7}
    & K=5 & K=10 & K=20 & K=5 & K=10 & K=20 \\
    \midrule
    GenPOI (w/o TCG) & 52.28 & 64.64 & 78.48 & 64.86 & 74.84 & 83.46 \\
    \textbf{GenPOI} & \textbf{0.00} & \textbf{0.00} & \textbf{0.00} & \textbf{0.00} & \textbf{0.00} & \textbf{0.00} \\
    \bottomrule
  \end{tabular}
\end{table}
We define the Invalid Generation Rate (IGR) as the proportion of generated PIDs that fail to map to a valid entity. As shown in Table~\ref{tab:invalid_rate}, without PID trie constrained generation (TCG), the model produces a high volume of invalid identifiers, with the IGR reaching up to 78.48\% on TMap-S and 83.46\% on TMap-L. This significant failure rate stems from the fact that our generation process involves a complex sequence of tokens representing both GID and SID. The vast combinatorial space of these tokens increases the probability of the model generating ``hallucinated'' sequences that do not exist in the database. However, by incorporating TCG, the IGR is consistently reduced to 0.00\% across all settings. By imposing hard constraints from the PID trie at every decoding step, TCG ensures that the generated token sequences strictly follow valid paths in the search space.

\section{Conclusion and Future Work}
In this paper, we propose GenPOI, an innovative generative Point-of-Interest retrieval framework specifically tailored for map search scenarios. By adopting a generative retrieval paradigm, GenPOI achieves an effective fusion of heterogeneous search contexts, leveraging the robust contextual modeling capabilities of LLMs to resolve complex and underspecified queries. Central to our approach is the geo-semantic POI tokenization module, which harmonizes POI attributes with geographic coordinates into structured, sequential identifiers (PIDs). To ensure operational reliability, we introduce a proximity-aware constrained generation strategy, guaranteeing both the validity and spatial relevance of the retrieved results. Extensive experiments on large-scale industrial datasets from Tencent Maps demonstrate that GenPOI consistently outperforms state-of-the-art baselines.

In future work, we plan to investigate contextual compression techniques to further enhance retrieval efficiency for real-time deployment, and integrate multi-modal features to enrich POI representations and further enhance retrieval performance.

\section*{Acknowledgment}
The authors would also like to thank the anonymous reviewers for their constructive comments and valuable suggestions.

\bibliographystyle{IEEEtran}
\bibliography{main}

\vfill

\end{document}